\documentstyle[aps,amssymb]{revtex}
\catcode`\"=\active \let"=\" \let\3=\ss

\begin{document}
\author{John H. Jefferson$^{1}$ and Wolfgang H\"{a}usler$^{2}$}
\address{$^{1}$DERA, Electronics Sector, St. Andrews Road, Malvern, \\
Worcs.~WR14\\
3PS,~United Kingdom \\
$^2$I.~Institut f\"ur Theoretische Physik, Jungiusstr.~9, 20355 Hamburg,\\
Federal ~Republic of~Germany}
\title{{\bf Quantum dots and artificial atoms}}
\date{\today}
\maketitle

\begin{abstract}
The electronic properties of nanoscale quantum dots are reviewed. The
similarities and differences between these `artificial atoms' and real atoms
are discussed and, in particular, the effect of electron correlations is
examined. It is shown that for (semiconductor) quantum dots, with only a few
electrons, electron correlations can give rise to important new physics
which is absent in their true atomic counterparts.
\end{abstract}

\footnotetext{$^2$ Present address~: Theoretical Physics Institute, 116
Church Str. SE Minneapolis MN 55455, U.~S.~A.}

\section{Introduction}

In 1985 R.P.Feynman made the following prophetic statement. ``{\em It seems
to me that the laws of physics present no barrier to reducing the size of
computers until bits are the size of atoms and quantum behaviour holds sway}
''\cite{feynman}. Well, here we are only a decade later and technological
advances in microfabrication have enabled us to not only manipulate
individual atoms but even single electrons. It is the manipulation of
single-electrons, and the properties of a few interacting electrons confined
on small islands of semiconducting material, called `quantum dots', that we
shall be concerned with here. Such quantum dots typically consist of a few
hundred to a few million atoms are thus still some way from being the single
atoms to which Feynman referred. Nevertheless, these quantum dot islands can
behave in many ways like single atoms as pointed out by Kastner, who coined
the term `artificial atom' \cite{kastner}. Certainly they can exhibit many
of the quantum effects associated with their true atomic counterparts, where
likewise few electrons are held together by the nuclear potential. However,
these structures exhibit new physics which, whilst being quantum in origin,
have no analogue in real atoms and influences their electrical properties.
These effects, which we will describe below, are due to the interactions
between the electrons. It turns out that {\em correlations} dominate the
physics in many cases and render the independent-electron approximation
useless, leading to even qualitatively incorrect results \cite{daniela93a}.
This should be compared with real atoms for which Hartree-Fock is an
excellent approximation, providing at least the correct qualitative
behaviour of atoms and a means for obtaining solutions of arbitrary accuracy
in a controlled way by configuration mixing and perturbation theory. The
origin of this different behaviour for artificial atoms with a small number
of electrons is the nature of the confining potential which, being usually
harmonic, is much more shallow than the $1/r$ potential of real atoms. This
results in a lower electron density (larger mean separation between
electrons) than is the case for atoms and the electron configurations can
approach the classical (Wigner) limit in which the electrons are in fixed
positions which minimise the electrostatic energy.

\section{Experimental realisation}

\label{experiment}

Single electron effects can be observed in the electrical transport
properties of small conducting particles that are weakly coupled to their
environment \cite{charging,singlecharge}. A simple example is a small
(nanoscale) metallic grain surrounded by a thin insulating layer. Indeed
such metallic `dots' can be fabricated in a controlled way these days, for
example gold particles encapsulated by organic molecules and placed on a
gold surface \cite{andres}. Electrons can be made to tunnel into the dot and
out through the substrate to produce a current using a fine probe such as an
STM tip. When this is done, a finite voltage is required at low temperatures
before any current is produced and the $I-V$ curve is antisymmetric about $%
V=0$, as shown in Fig. 1. This finite turn-on voltage is referred to as the
`Coulomb gap' and occurs because of the very small capacitance of the
island, as we explain in the next section. An even more spectacular effect
is observed when a third (gate) electrode is introduced, as we describe
below. A schematic diagram of one such structure, similar to that fabricated
by Meirav {\it et al} \cite{meirav91}, is shown in Fig. 2. It consists of a
heterojunction between two semiconductors (such as GaAs and AlAs), a top and
bottom gate (the top gate being patterned) and source and drain contacts.
The wider gap semiconductor is doped with donors and the carriers migrate to
the narrower gap semiconductor where they form a thin sheet of charge at the
heterojunction interface, {\it i.e.} a quasi two-dimensional electron sheet
(2DES). When the top gate is biased negatively, the electrons beneath it in
the 2DES are repelled and, without the four `stubs' on the gate, this would
give rise to a thin filament of charge, converting the 2DES into a quasi
one-dimensional wire. The purpose of the stubs is to give rise to barrier
for electrons, which defines a small potential well, as shown in the
electron potential energy diagram of Fig. 3. (This is for a plane close to
the heterojunction interface where the charge density would be a maximum for
the 2DES in the absence of the top gate.) The Fermi energy of the electrons
near the heterojunction interface can be adjusted by changing the voltage on
one or both gates. At low temperatures there is little or no current through
the device when a small source-drain voltage is applied {\em except} in the
region of certain gate voltages when a current flows. This behaviour is
shown in the $I-V_{G}$ curve of Fig. 4. The current switches on and off as
the gate voltage is increased. This behaviour can be explained roughly as
follows (a more precise explanation will be given in the next section). The
gate voltages may be adjusted such that the Fermi energy lies below the
bottom of the quantum well but still above the quasi 1D `leads' at either
side of the well as shown in Fig. 5a, where the shaded region represents the
quasi 1D Fermi sea. The forbidden gap region below the well is too wide to
permit tunnelling in one jump from one lead to the other and hence at low
temperature there is negligible current. If the gate voltage is adjusted so
that the quantum well is lowered in energy (by making the bottom gate
voltage more positive) then eventually the Fermi energy will be above the
bottom of the well and a current will flow with electrons tunnelling into
and out of the dot (Fig. 5b). This is rather like what happens in a field
effect transistor (FET), which is `switched on' by applying a gate voltage.
However, unlike the FET, the device may be switched off by further
increasing the gate voltage. This is because there are no empty states in
the leads available for the electron to leave the dot when the well is
lowered in energy. Furthermore, a second electron may also not enter the dot
because of the large energy associated with the Coulomb repulsion between
two electrons (or, equivalently, the very tiny effective capacitance of the
dot). This switching off of the current is known as {\em Coulomb blockade}$.$
A further increase in the gate voltage lowers the energy of the electron on
the dot sufficiently that a second electron may enter the dot. This occurs
when $E_{F}=E_{0}+U$, where $E_{F}$ is the Fermi energy in the lead, $E_{0}$
is the lowest energy in the well and $U$ is the Coulomb repulsion energy
between the two electrons on the dot. Since empty states are also available
at energy $E_{F}$ in the drain lead, an electron may leave the dot again,
giving rise once more to a finite current. In this way we get a sequence of
sharp spikes in the current vs gate voltage curve as shown in Fig. 4. We
also point out that the Coulomb blockade is still effective when there is a
current flowing. The number of electrons on the quantum dot can be either $N$
or $N+1$ but other values are suppressed. An electron hopping onto the dot
must leave it before another electron may enter. In this way the current is
due to single electrons with mean spacing determined by the tunnelling time
through the strongest barrier.

The quantum dots described above have some similarity with atoms in the
sense that they usually have an integral number of electrons (Coulomb
blockade mode) and there is an energy cost to either remove an electron
(ionisation) or add an electron (affinity), as with atoms. Furthermore, the
number of electrons on the dot may be changed by changing the Fermi energy.
This is analogous to a change of valence of real atoms which may be effected
by changing the local chemical potential. The analogy with atoms is somewhat
closer for semiconductor dots than for metallic grains since the latter will
typically contain many millions of conduction electrons whilst the former
will usually have a small number, with $N=0,1,2,\ldots $ corresponding to
artificial $H^{+},H,He,\ldots $ \cite{dots}. However, despite these
similarities, there are important differences between real atoms and
artificial atoms as we shall see in section \ref{energy}.

\section{Charging model}

\label{charging model}

There is a simple semi-classical model which explains with reasonable
accuracy the behaviour of artificial atoms as described in the previous
section \cite{charging}. Consider a metallic particle on an insulating
surface to which we attach source leads via thin insulating (tunnel)
barriers, as shown in Fig. 6. We introduce a positively charged `gate'
electrode close to the metal particle but sufficiently insulated from it
that charge may may not tunnel. By applying a small bias between the leads
we observe that the current shows oscillations with gate potential, similar
to those of Fig. 4. A semi-quantitative explanation is as follows. Let the
capacitance of the metal particle be $C$ and the excess charge on it be $Q$.
If the potential on the metal particle is $V$ then the electrostatic energy
is 
\begin{equation}
E_{Q}=QV+\frac{Q^{2}}{2C}  \label{energyQ}
\end{equation}
where $Q=-Ne$ for an excess $N$ of electrons. As $V$ is increased, electrons
will flow from the leads onto the dot, where the potential is lower (first
term in Eq. (\ref{energyQ}) ). However, opposing this is the large Coulomb
repulsion between charges on the dot, represented by the small capacitance $%
C $ (second term). At absolute zero, the total energy will be a minimum
which, for this parabolic equation, gives $Q=-CV$. However, this is not
quite correct since $Q$ is not a continuous variable: the quantum dot must
contain an integral number of electrons. $N$ is thus the integer which makes 
$E_{-Ne} $ closest to $E_{-CV}.$ This is shown in Fig. 7a, where we also
show the total energy with one extra electron and one fewer electron,
corresponding to electron affinity and ionisation respectively. Denoting
these energies by $E_{A}$ and $E_{I}$, it follows directly from Eqn. (\ref
{energyQ}) that 
\[
E_{A}+E_{I}=\frac{e^{2}}{C} 
\]
Furthermore, at $T=0$ the current will be blocked for a very small potential
difference between the leads due to the electron affinity and ionisation
barriers. This is the Coulomb blockade which can, of course, be overcome at
finite temperatures. Its effect is maximal when $E_{A}=E_{I}=\frac{e^{2}}{2C}
$ , as shown in Fig.7b, and minimal when there are two dot occupation
numbers with exactly the same energy, as shown in Fig. 7c. For this latter
case, there will be a ``single-electron current'' even at absolute zero.
Hence, this simple model predicts current or conductance oscillations with
equidistant peaks separated by $\Delta V=e/C$, which agrees approximately
(though not precisely) with experiment. The model also explains the
``Coulomb gap'' shown in Fig. 1, which occurs when the gate voltage is held
constant but the voltage between the leads increased. Consider the arbitrary
case shown in Fig. 7a. At absolute zero there will be no current for a
potential difference less than $E_{A}/e$. If the potential on the left lead
(say) is increased ({\it i.e.} the potential energy of electrons is
lowered), then an electron in the dot will eventually have the same energy
as the Fermi energy in the left lead and a one-electron current will flow
from right to left. This will occur when the potential in the left lead has
been raised by $E_{I}/e$. On the other hand, starting again with zero
potential difference between the leads, if the potential on the right lead
is decreased, an electron at the Fermi energy will be able to transfer to
the dot with no energy penalty when the potential is $-E_{A}/e$ and a
current will flow from right to left. Hence the potential difference between
the onsets of left and right currents is 
\[
\Delta V=\frac{E_{I}+E_{A}}{e}=\frac{e}{C} 
\]
This relation can, of course, be used to estimate the effective capacitance
in a two-terminal experiment. For observation at $4.2K$ or higher
temperatures, this effective capacitance has to be very small, $\lessapprox
10^{-17}F$ .

\section{Energy spectrum of the artificial atom - correlations}

\label{energy}

The semiclassical charging model seems to provide a simple explanation of
the main features of quantum dots and their similarity with atoms so do we
only need to refine this theory to improve quantitatively the agreement with
experiment? How far can we push the analogy of quantum dots as artificial
atoms? Do they share other features of real atoms, such as a discrete energy
spectrum, shell structure, Hund's rules etc.? There is now significant
evidence, both theoretical and experimental, to support the view that there
are some fundamental differences between real and artificial atoms. A simple
analysis of the length and energy scales involved indicate that this might
be the case. Within the parabolic-band effective-mass approximation, the
natural units of energy and length are the scaled Rydberg and Bohr radius: 
\[
R_{y}^{*}=\frac{m^{*}}{m_{0}\varepsilon ^{2}}R_{y},\;a^{*}=\frac{\varepsilon
m_{0}}{m^{*}}a 
\]
These would be, for example, the ground-state binding energy and
corresponding Bohr radius of a conduction electron bound to a shallow donor
ion in a semiconductor. We show below, a table of these quantities for the
semiconductors Si, GaAs and InSb, together with H for comparison. 
\[
\begin{tabular}{|c||c|c|c|c|}
\hline
& H & Si & GaAs & InSb \\ \hline\hline
$R_{y}^{*}({\rm meV})$ & $13,600$ & $18$ & $8$ & $0.85$ \\ \hline
$a^{*}({\rm A})$ & $0.529$ & $33.4$ & $82.4$ & $567$ \\ \hline
\end{tabular}
\]
We see that the energies are typically $\sim 1000$ times less than H with
Bohr radii $\sim 100$ times greater! Furthermore, both with real atoms and
shallow impurities in semiconductors, the length scale is dictated by the
Coulomb attraction of the nucleus and the orbiting electrons. For atoms with
more than one electron, this ensures that the kinetic energy remains greater
than Coulomb repulsion energy between electrons. For example, in He the mean
kinetic energy is almost three times the mean Coulomb repulsion energy
between the two $1s$ electrons. For this reason, the independent electron
approximation works very well for atoms. Hartree-Fock gives qualitatively
the correct picture and even quantitatively errors are only about 5\% for
light atoms. For quantum dots, however, the situation can be quite
different. Firstly the positive potential which confines electrons is
usually much more gradual than the Coulomb potential of the atomic nucleus
on the scale of the electron wavelength. For large dots, such as those
produced by `soft' confinement as in systems fabricated on the basis of
semiconductor heterojunctions, this will generally be parabolic. For `hard'
confinement, such as is produced by a heterojunction between two dissimilar
semiconductors, the potential can be approximately constant inside the dot.
Let us for simplicity consider two electrons in a quantum box with sides of
length $L$ and infinite barriers. The one-electron level spacing at
low-energies due to kinetic energy is, 
\[
\Delta E_{K}\sim \frac{\hbar ^{2}}{m^{*}L^{2}} 
\]
independent of dimensionality. On the other hand, the Coulomb repulsion
energy between the two electrons is 
\[
E_{C}\sim \frac{e^{2}}{\varepsilon \varepsilon _{0}L} 
\]
Hence 
\[
\frac{\Delta E_{K}}{E_{C}}\sim \frac{a^{*}}{L} 
\]
where we have used $a^{*}=\frac{\varepsilon \varepsilon _{0}\hbar ^{2}}{%
m^{*}e^{2}}$. This shows that for sufficiently large dots and given electron
number, the Coulomb repulsion energy between electrons will dominate. For
example, in a GaAs quantum dot, Coulomb effects are expected to dominate for 
$L>0.15$ microns whereas in Si this would be about $500{\rm A}.$ \footnote{%
This argument is a little too simplistic. For high confinement
non-parabolicity effects in k-space and the finite height of the confining
barriers can have a significant effect. Nevertheless these estimates are
expected to be reasonable lower limits for the critical length scales beyond
which Coulomb interactions dominate the physics.}. We will show in the next
section how the relatively large Coulomb energy is expected to give rise to
qualitatively different physics. This yields an energy spectrum for the
artificial atom which is quite different from that given by the independent
electron approximation, due to the correlations between electrons. However,
in the very first experiments on small dots, where the excitation energies
are expected to be sufficiently large to be observable, no deviation from
the independent-electron picture was detected by means of (artificial)
atomic (far infrared) spectroscopy \cite{fir}. The resolution of this
paradox lies in a theorem due to Kohn \cite{Kohn}, which states that the
optical spectrum of an interacting electron system which is confined by a
parabolic potential is independent of the interactions between electrons for
dipole transitions. This is because the radiation field couples only to the
centre-of-mass co-ordinate of all the electrons, the motion of which does
not depend on electron-electron interactions. Although, strictly speaking,
the confining potentials in quantum dots will not be perfectly parabolic, it
turns out in practise that deviations from parabolicity are on the
borderline of being observable which makes simple dipole excitations less
favourite for probing correlation effects.

The other class of experiments are transport measurements of, for example,
electrons tunnelling through a quantum dot in a single-electron transistor 
\cite{meirav91,leo91,heinzel94a}. However, for the experiments described in
Section \ref{experiment}, which show oscillations in current vs gate voltage
for very small source-drain voltage, we do not expect to see a dramatic
effect of electron correlations either. This is because such experiments
probe only the interacting ground-states of the quantum dot. The difference
in ground-state energy when the number of electrons on the dot is changed by
one in not very small quantum dots is accounted for with reasonable accuracy
by the Coulomb charging model, which predicts a constant separation between
peaks in the conductance oscillations. Nevertheless, in very recent
experiments, where it has become possible to fabricate quantum dots
containing down to $N=1,2,\ldots $ electrons and to contact them for
transport experiments, significant deviations from the predictions of the
simple model have been found, particularly for the first few peaks\cite
{tarucha96,sivan,wharam,zhitenenv}. These deviations can be attributed to
correlations. An even more significant probe would be to look at excited
states. This can be done by simply increasing the source-drain voltage. In
fact, by doing so we may regard the single-electron transistor as an {\em %
energy spectrometer} for quantum dots/artificial atoms! That such an
experiment can probe the excited states of the quantum dot can be understood
from the following argument, which closely follows that given at the end of
Section \ref{charging model} to explain the Coulomb gap. Suppose that at low
temperature we adjust the gate voltage of the single-electron transistor
just between two conductance oscillation peaks. Let there be $N$ electrons
on the dot. Suppose that the Fermi energy of the drain contact is now
lowered by changing its voltage. Eventually the ground-state energy of the $%
N $ electrons will be the same as the ground state energy of $N-1$ electrons
on the dot plus an electron at the Fermi energy in the drain lead. This will
cause a current to flow as an electron may leave the dot allowing another
electron to enter it from the source lead. However, the {\em magnitude} of
this current will depend strongly on the position of the Fermi energy in the
source contact. If this is such that $E_{F}^{{\rm source}%
}+E_{N-1}^{0}<E_{N}^{1}$, where $E_{N-1}^{0}$ is the ground-state energy of
the $N-1$ electrons on the dot and $E_{N}^{1}$ is the first excited state of 
$N$ electrons on the dot, then the current will be due (only) to electrons
in the source at energy $E_{N}^{0}$ $-E_{N-1}^{0}$ tunnelling into the dot.
However, when $E_{F}^{{\rm source}}+E_{N-1}^{0}=E_{N}^{1}$ there are now two
channels for tunnelling into the dot, {\it i.e.} electrons at the Fermi
energy and electrons at energy $E_{N}^{0}$ $-E_{N-1}^{0}$. Hence, at this
point there will be a sudden rise in current. These steps of increasing
current occur whenever $E_{F}^{{\rm source}}+E_{N-1}^{0}=E_{N}^{n}$, where $%
E_{N}^{n}$ is an excited state with $N$ electrons. Hence there will be a
fine structure at either side of the Coulomb gap reflecting the energy level
structure of the dot \cite{averin91,beenakker91}. This has indeed been
observed in experiments that show a rich structure of resonance steps
corresponding to excited states \cite{trspectra}. They also show some
unusual features which cannot be explained by an independent-electron
picture, such {\em negative differential resistance}. As we shall see in the
next section, this kind of unusual behaviour is indeed expected for strongly
correlated electrons on the quantum dot.

\section{Theory}

\label{theory}

It is well known that electrons interacting via Coulomb repulsion in a
uniform background of positive charge behave quite differently in the high
and low density limits. At relatively high densities comparable with that in
real metals, the kinetic energy dominates and, despite the still large
Coulomb repulsion, the system at low-energies may be mapped onto an
independent-electron model of weakly interacting quasiparticles.\cite{landau}
At low-densities this picture is not even qualitatively correct. The
potential energy (Coulomb repulsion) dominates over kinetic energy and the
electrons eventually form a lattice \cite{wigner34}, with low-lying
excitations that are lattice vibrations \cite{chaplik72,meissner76b,tsidil87}
(see also the article by Ando, Fowler and Stern \cite{ando82}). This occurs
for a mean separation between neighbouring electrons $\gtrsim 100a_{{\rm B}}$%
, where $a_{{\rm B}}$ is the effective Bohr radius (denoted by $a^{*}$ in
the previous section). For intermediate densities, this is an extremely
difficult problem due to competing interactions from the quasi-classical
`lattice' vibrations and quantum mechanical exchange. Physical realisation
of the Wigner lattice is an extremely challenging experimental problem due
to the difficulty in obtaining a low-defect region with sufficiently low
electron density. Nevertheless evidence of Wigner crystallisation has been
reported in heterostructures \cite{timofeev92,clark91}.

With large quantum dots, we have the interesting possibility of the analogue
of Wigner crystallisation for finite systems with just a few electrons \cite
{peeters94}. In contrast to the infinite system, not only might the
density--density correlation function crystallize but also the charge
density distribution itself, with the electrons occupying energetically
favourable positions within the dot. This can indeed be the case as we shall
see, with the geometry of the confining region being crucial to the nature
of the few-electron states in two and three dimensions. By analogy with the
Wigner lattice, these quantum dot systems with large mean separation between
electrons have been called Wigner molecules \cite{maksym93,jauregui93}. One
theoretical advantage of studying few-electron systems is that they have the
prospect of essentially being solved exactly by direct numerical
diagonalisation of the Hamiltonian matrix \cite{numerics}. It turns out that
they are also amenable to approximations which are not valid for large
systems (see below). Such approximations are still necessary, for although
we are interested in a few-electron problem (with $N$ $\lesssim 20$), it is
prohibitively expensive on computer time to solve for more than a very few
electrons by direct diagonalisation. For sufficient accuracy, the practical
limit is $N=5$ in one dimension and $N=3$ in two dimensions. For high
density systems we can, of course, get reasonable results with the
independent-electron model using the Hartree-Fock approximation, for which
solutions with tens of electrons are easily obtained \cite{gudmundsson}.
However, this is a poor approximation when the mean electron separation is
large and gives qualitatively incorrect results.

This may be seen from exact diagonalisations for a one-dimensional quantum
dot with a simple rectangular well confining potential \cite
{haus93,jauregui93}. We summarise the main results of these computations in
Figs. 8 and 9. Figure 8 shows the energy spectrum for $N=1$ to $4$ for a
well width $L=9.45a_{{\rm B}}$. This is seen to consist of a series of
multiplets for which the energy separation between multiplets is much
greater than the energy splitting within a multiplet. Furthermore we can
easily see that this spectrum is qualitatively different from that which
would be given by an independent-electron picture by examining the lowest
multiplet with $N=3$ electrons. In the independent-electron picture, we
expect a doublet lowest followed further doublets at energies $\varepsilon
_{1}-\varepsilon _{0}$ and $\varepsilon _{2}-\varepsilon _{1}$ (relative to
the ground-state) and an octet at energy $\varepsilon _{2}-\varepsilon _{0}$
etc., where $\varepsilon _{n}$ are the one-electron levels. Instead we see a
low-lying multiplet consisting of doublet, doublet, quartet with very small
splitting between them ($\ll \varepsilon _{1}-\varepsilon _{0}$), followed
by a (relatively) very large gap to another multiplet of closely spaced
levels. A clue as to what is causing this level structure is found in Fig. 9
where the charge density due to all three electrons in the ground-state of
the interacting system is plotted for various $L$. These curves are
qualitatively different for large and small $L$. For $L=0.1a_{{\rm B}}$, we
get a charge density characteristic of the independent-electron picture,
since this would place two electrons in the nodeless ground-state and the
third electron in the first excited state with a node at $L/2$, giving a
shallow minimum in the total electron density. At the other extreme, with $%
L=945a_{B}$, we see three distinct peaks. This is close to what we would
expect in the `Wigner limit', where the electrons arrange themselves with
equal separation in order to minimise the electrostatic energy. This
quasi-classical configuration is essentially spin-independent, as can be
seen from the corresponding ground-state multiplet which is almost (but not
quite) degenerate. These results are not unexpected. What is less obvious is
the behaviour at smaller $L$, where we see a persistence of the Wigner
picture down to rather small $L$, the three peaks being discernable even
when $L=1.9a_{{\rm B}}$ ({\it i.e.} mean electron separation of order $a_{%
{\rm B}}$)! Furthermore, an examination of the energy spectrum shows that
the ground-manifold spin-multiplet, with $2^{3}=8$ states, remains well
separated from higher-lying multiplets as $L$ is reduced. The main effect is
to increase rapidly the magnitude of the splitting within the ground-state
multiplet as $L$ is decreased whilst (approximately) preserving their ratio
of order $2:1$. We see a similar behaviour for $N=4$ with a ground-manifold
spin-multiplet of dimension $2^{4}=16$, which splits into two singlets,
three triplets and a quintet.

In the infinite system the low-lying excitations are sound waves in the
Wigner limit when the spin can be ignored, since the electrons become
distinguishable through their respective lattice sites. Indeed, this is what
makes the problem prohibitively difficult at higher densities where the
Fermi statistics of the electrons can no longer be ignored. The major
simplification of the few-electron system is that these sound-wave-like
excitations are high in energy compared to low-energy spin-splittings and
can therefore be safely ignored \cite{haus95}, thus permitting the effect of
the Fermi statistics to be examined in isolation as $L$ decreases. This is
similar to the situation in real molecules, where the separation between
vibrational levels is greater than the fine-structure due to electron
interactions. (Whereas in crystalline solids both lattice vibrations and
electronic excitations are present in the low-energy spectrum.)

In the remainder of this section we will concentrate mainly on the
low-energy manifold of states for quantum dots in both one and two
dimensions. Consider again the simplest non-trivial case of two electrons in
a one-dimensional rectangular well, interacting via a potential of the form 
\begin{equation}
V(|x_{1}-x_{2}|)=\frac{e^{2}}{\varepsilon \varepsilon _{0}\sqrt{%
(x_{1}-x_{2})^{2}+\lambda ^{2}}}  \label{V2}
\end{equation}
This has the usual Coulomb form for large separations between the electrons
and passes through a maximum when the $x$--coordinates coincide. It models a
quasi-1D quantum dot, the parameter $\lambda $ representing the confinement
in the other two dimensions, with $\lambda $ decreasing as the confinement
increases. We may regard this two-electron problem in 1D as a
single-particle problem in 2D with momentum {\bf p}=($p_{1},p_{2}$) and
position {\bf r}=($x_{1},x_{2}$). The potential landscape for this
fictitious particle, $V(r),$ is shown in Fig. 10 where we see potential
minima at ${\bf r}=(L,0)$ and $(0,L)$, corresponding to the two positions
where the `real' particles (electrons) are maximally separated. When viewed
in this way, the problem is reminiscent of a tunnelling problem in which the
particle may start in one well, or `pocket' and will subsequently tunnel to
the other. To describe this situation quantitatively we may choose basis
states for the particle in each well. These may be constructed, for example,
by replacing the potential energy in region 2 of Fig.~10 by the maximum
value $V_{0}\equiv V(0)=\frac{e^{2}}{\varepsilon \varepsilon _{0}\lambda }$
and solving for the eigenstates in region 1, $\psi _{1}^{(n)}$ say.
Similarly for the well states in region 2, $\psi _{2}^{(n)}$. Now, if the
barrier of the true potential (Fig.~10) is sufficiently strong then, to a
good approximation, there will be two low-lying states 
\[
\psi _{\pm }=\frac{(\psi _{1}^{(0)}\pm \psi _{2}^{(0)})}{\sqrt{2(1\pm s)%
\text{ }}} 
\]
with corresponding energies 
\[
E_{\pm }=\frac{E_{0}\pm t}{1\pm s} 
\]
where $E_{0}=\langle \psi _{1}^{(0)}|H|\psi _{1}^{(0)}\rangle =\langle \psi
_{2}^{(0)}|H|\psi _{2}^{(0)}\rangle $, $t=\langle \psi _{1}^{(0)}|H|\psi
_{2}^{(0)}\rangle $ and $s=\langle \psi _{1}^{(0)}|\psi _{2}^{(0)}\rangle $.
Furthermore, these eigenstates will be well separated from higher-lying
states. To complete the picture, we must impose the indistinguishability of
the underlying electrons on these solutions. This is straightforward since
we know that the symmetric solution, $\psi _{+}$, corresponds to a spin
singlet and the antisymmetric solution, $\psi _{-}$, to a spin triplet.

The important point about the above analysis is that we may deduce the
nature of the low-lying states without explicitly solving the problem,
provided we can justify the neglect higher lying states in each well. This
is indeed the case for the potential given in Eqn.\ref{V2}. It is emphasised
that the base states themselves are highly correlated and cannot in general
be approximated by a simple product of one-electron wavefunctions.

We may extend this `pocket state' analysis to cases with more electrons and
higher dimensions. Thus $N$ electrons in $D$ dimensions is equivalent to a
single particle in a $N\times D$ dimensional space. This fictitious particle
will move in a potential landscape with at least $N!$ minima corresponding
to the classical minimum electrostatic energy of the $N$ electrons regarded
as distinguishable (though equivalent). In two and three dimensions there
may be more minima for certain geometries. For example, consider 3 electrons
in a square dot. The electrons will be repelled to the corners of the square
and there will be four different configurations corresponding to one corner
unoccupied and the remaining 3 occupied. Hence there will be $4\times 3!=24$
classical minima. In the general case there will be $\nu N!$ equivalent
global minima. Provided these are well separated in energy from possible
higher local minima then the structure of the low-lying eigenstates may
again be deduced from the Hamiltonian matrix of the system within the basis
of the $\nu N!$ pocket states of lowest energy. For $N>2$ the requirement
that the true eigenstates be antisymmetric with respect to interchange of
both position and spin is not trivial since, unlike the case of $N=2$, the
`orbital' part of the wavefunction will be neither symmetric nor
antisymmetric. However, the problem may be solved using the theory of
permutation groups and the corresponding low-lying spectrum deduced \cite
{haus94,haus95}.

The pocket state analysis suggests an alternative approach to the
few-electron quantum dot problem which is similar to that used for strongly
correlated lattice systems. The generic theoretical model for such lattice
problems is the Hubbard model. This has been successful in modelling Mott
insulators, the metal-insulator transition and, more recently, doped Mott
insulators, which are believed to be relevant to the copper-oxide planes of
high-temperature superconductors in which the physics is believed to be
dominated by electron-electron correlations.\cite{dagotto} We can indeed
show that this model is relevant to interacting electrons in a quantum dot
but let us first review the main properties of the Hubbard model.

The Hubbard model is essentially a tight-binding model in which only the
largest Coulomb matrix elements are retained. Consider, for example, a
collection of one-electron atoms which are essentially in fixed positions
but otherwise quite general (molecule, crystalline solid or disordered
solid). Following Hubbard \cite{hubbard1} we approximate this system by
confining the Hilbert space to just one orbital per atom and drop all
Coulomb matrix elements except the largest, which corresponds to two
electrons of opposite spin on the same site (atom). The resulting
Hamiltonian is 
\begin{equation}
H=\sum\limits_{i}\left( \varepsilon _{i}n_{i}+U_{i}n_{i\uparrow
}n_{i\downarrow }\right) +\sum\limits_{\langle ij\rangle \sigma }\left(
t_{ij}c_{i\sigma }^{\dagger }c_{j\sigma }^{{}}+{\rm H.c.}\right)
\label{hubbard}
\end{equation}
where {}$n_{i}=n_{i\uparrow }+n_{i\downarrow }$, $n_{i\sigma }=c_{i\sigma
}^{\dagger }c_{i\sigma }^{{}}$, $\varepsilon _{i}$ is a one-electron energy
for atom $i$, $t_{ij}$ is a hopping matrix element from atom $i$ to atom $j$
and $U_{i}$ is a Coulomb matrix element for two electrons on atom $i$. In
this equation, the angular brackets in the summation over $i$ and $j$
denotes pairs of atomic sites in which each pair is counted only once. The
one-electron wavefunction on atom $i$ is $\psi _{i}({\bf r})\chi _{\sigma
}\equiv \langle {\bf r|}c_{i\sigma }^{\dagger }|{\rm vac}\rangle $ and 
\[
\varepsilon _{i}=\langle {\psi _{i}|}\left[ \frac{{\bf p}^{2}}{2m}+v({\bf r}%
)\right] |{\psi _{i}\rangle ,} 
\]
\[
t_{ij}=\langle {\psi _{i}|}\left[ \frac{{\bf p}^{2}}{2m}+v({\bf r})\right] |{%
\psi _{j}\rangle ,} 
\]
\[
U_{i}=\dot{\int }{|\psi _{i}({\bf r})|^{2}|\psi _{i}({\bf r}}^{\prime }{%
)|^{2}}\frac{{e}^{2}}{|{\bf r}-{\bf r}^{\prime }|}{{\rm d}^{3}r{\rm d}^{3}r}%
^{\prime } 
\]
where $v({\bf r})$ is some (optimal) one-electron pseudopotential. In some
situations (for example when there is significant charge transfer from one
atom to another) the Coulomb interaction between electrons on different
atoms can be important and we have to add the term 
\begin{equation}
\sum\limits_{\langle ij\rangle }V_{ij}n_{i}n_{j}  \label{vij}
\end{equation}
to Eq. (\ref{hubbard}), where 
\[
V_{ij}=\dot{\int }{|\psi _{i}({\bf r})|^{2}|\psi _{j}({\bf r}}^{\prime }{%
)|^{2}}\frac{{e}^{2}}{|{\bf r}-{\bf r}^{\prime }|}{{\rm d}^{3}r{\rm d}^{3}r}%
^{\prime } 
\]
The resulting Hamiltonian is sometimes referred to as the extended Hubbard
model. When the separation between the atoms is sufficiently large, which is
the regime of interest to us here, it follows that the $t\ll U$, for all $t$
and $U$. This is known as the strongly correlated regime since the motion of
the electrons are strongly dependent on each other and is truly a many-body
problem. (The amplitude for an electron to hop from one atom to another is
strongly dependent on whether the recipient atom is occupied or not because
of the large $U$.) Despite this being a difficult problem, a certain
simplification arises in this strong correlation regime as many of the
states in the many-electron Hilbert space may be eliminated by perturbation
theory in $t/U$ or, equivalently, a canonical transformation.\cite
{lindgren,jjlett,SW,chao}

For the important case in which the number of electrons equals the number of
atoms, it may be shown that the low-energy states of $H$ are given by an
effective spin model in which each atom always has one electron but
electrons (spins) on neighbouring sites interact via an antiferromagnetic
Heisenberg exchange interaction, {\it i.e.} the effective Hamiltonian 
\begin{equation}
H_{{\rm eff}}^{{}}=\sum_{\langle ij\rangle }J_{ij}{\bf s}_{i}\cdot {\bf s}%
_{j}  \label{heisenberg}
\end{equation}
has, apart from an unimportant constant energy shift, the same low-energy
spectrum (to second-order) as the original Hamiltonian, $H$. In this
equation the positive exchange interactions $J_{ij}$, often referred to as
superexchange, are given by 
\[
J_{ij}=2|t_{ij}|^{2}\left[ \frac{1}{U_{i}}+\frac{1}{U_{j}}\right] 
\]
(This gives a total exchange for a pair of atoms of $J{\bf s}_{i}\cdot {\bf s%
}_{j}$ where $J=4|t|^{2}/U$ for a homogenous system.) This reduction to a
Heisenberg spin model simplifies the Hubbard models in two ways. Firstly, it
drastically reduces the size of the Hilbert space since each site can be in
one of only two states compared with four for the Hubbard model. Secondly,
it gives a new language and insight to describe the low-energy spectrum.

When the number of electrons does not equal the number of atoms this means
some sites must be unoccupied ($N_{e}<N_{s}$) or doubly occupied ($%
N_{e}>N_{s}$), even in the low-energy manifold. In fact, these two regimes
are related to each other by electron-hole symmetry so we need only consider
one, $N_{e}<N_{s}$ say. For this regime states involving double occupations
are high in energy and, as with the case $N_{e}=N_{s}$ above, may be
eliminated by perturbation theory leading to an effective Hamiltonian for
the low-energy manifold: 
\begin{equation}
H_{{\rm eff}}^{tJV}=P\sum_{\langle ij\rangle }\left[ \sum_{\sigma
}(t_{ij}c_{i\sigma }^{\dagger }c_{j\sigma }^{{}}+{\rm H.c.})+J_{ij}({\bf s}%
_{i}\cdot {\bf s}_{j}-1/4)n_{i}n_{j}+V_{ij}n_{i}n_{j}\right] P
\label{tJVmodel}
\end{equation}
In this effective Hamiltonian the operator $P$ is a projection operator
which precludes all many-electron states in which sites are doubly occupied.
In the strong correlation regime, $t\gg J$, since $J\backsim t^{2}/U$ and $%
U\gg t$, and hence we may drop the $J$-term in (\ref{tJVmodel}) if we
neglect energy separations on the scale of $J$. The resulting $t-$model
would be a simple one-electron tight-binding model were it not for the $P$%
-operators. These make the model much more difficult to solve and can give
rise to unusual effects related to correlation implicit in $P$. For example,
a homogeneous system (lattice) with one `hole' ({\em i.e.} $N_{s}-1$
electrons) has a ground state of maximum spin.\cite{nagaoka}

To relate the low-density interacting-electron problem in a quantum dot to
the Hubbard model we start by approximating the pocket states by products of
one-electron states, $\psi _{i}({\bf r})$, centred about `sites' $i$ that
define the electrostatic ground-state energy. This is not an unreasonable
approximation provided that the one-electron states are chosen optimally,
which requires them to be non-orthogonal. With such a choice we could form a
set of antisymmetrized basis functions (Slater determinants) for the $N$
electrons by forming spin-orbitals, $\psi _{i}({\bf r})\chi _{i\sigma }$,
and antisymmetrizing products of $N$ such spin-orbitals, with the constraint
that no more than one electron shall occupy one `site'. It may be shown that
there is precisely a one-to-one mapping between these base states and
corresponding permutation-symmetry adapted pocket states, the total number
being $2^{N}$ for the (usual) case when the number of electrons equals the
number of `sites'.\cite{haus95} (An exception to this is considered below.)
However, such a procedure is complicated by the non-orthogonality of the
one-electron states. We can circumvent this problem and actually improve on
the approximation by orthogonalising the one-electron orbitals and expanding
the Hilbert space to allow any site to be occupied by either zero, one or
two electrons (subject, of course, to the constraint that the total number
of electrons be $N$). If we now express the Hamiltonian of the quantum dot (%
{\em cf} Eqn. (\ref{V2}))
\[
H_{{\rm dot}}=\sum_{i=1}^{N}\left[ \frac{{\bf p}_{i}^{2}}{2m^{*}}+v({\bf x}%
_{i})\right] +\sum_{\langle i,j\rangle }\frac{e^{2}}{\varepsilon \varepsilon
_{0}\sqrt{({\bf x}_{i}-{\bf x}_{j})^{2}+\lambda ^{2}}} 
\]
in this expanded basis set of Slater determinants we get, in second
quantized form, the Hubbard model Eq. (\ref{hubbard}), including the
intersite interaction Eq. (\ref{vij}).\cite{jeffhaus}

Having established this equivalence of the quantum dot problem to a Hubbard
model we may deduce properties of the low-energy spectrum for various cases.
In one-dimension we can transform the Hubbard model into a Heisenberg model
for the quantum dot. This is because the electrons will always arrange
themselves to form a `lattice' (chain) in which the number of lattice sites
equals the number of electrons. The lattice sites will, of course, be
determined by the electrons themselves. For example, in a rectangular well
of width $L$ the lattice will be uniform with spacings $L/(N-1)$ in the
Wigner limit. This should be contrasted with a chain of atoms for which the
number of lattice sites and their spacings are independent of electron
number. We have calculated the low-energy spectrum of one-dimensional
quantum dots with up to six electrons using the Heisenberg model and
confirmed that the results agree precisely with the pocket state method and,
where applicable, exact results. This also explains the nature of the lowest
spin-multiplets obtained by direct diagonalisation and described earlier in
this section. In all cases in one dimension, the ground-state is always
low-spin in accordance with the Lieb-Mattis theorem.\cite{liebmattis}

The equivalence between interacting electrons in a quantum dot and a Hubbard
model has important and interesting consequences in higher dimensions. These
arise for two reasons: (i) the number of `sites' may exceed the number of
electrons and (ii) the ground state need not necessarily be low-spin. We
will now consider these in more detail.

Case (i) is the analogue of the doped Mott insulator for lattice models and
is described by the $tJV$-model, Eq. (\ref{tJVmodel}). For the quantum dot,
the exchange parameters $J_{ij}$ are relatively very small and on the scale
of $t$ and $V$ they may be dropped. In some cases the $V$ may also be
dropped since it merely gives rise to an overall energy shift. For example,
three electrons in a two dimensional quantum dot of square symmetry. There
are four positions for the `hole' in this system, as shown in Fig. 11, and
all have the same contribution from the $V$-term. Dropping these and the $J$%
-term in Eq. (\ref{tJVmodel}) we are left only with the $t$-term, like a
simple tight-binding model. However, as stated earlier the presence of the
projection operators, $P$, have an important effect and the analytic
solution of the problem shows that the ground state is one of high-spin, 
{\it i.e.} a spin polarised state (see Fig.12a). This is purely an
interference effect in that the hole may gain maximum kinetic energy when
all spins are aligned and is the analogue of the spin-polarised state for a
single hole in a lattice system. \cite{nagaoka}.

A situation where the $V$-term {\em is} important is that of two electrons
in a square well in two dimensions, as shown in Fig. 13. The first two
states are lower in energy by an amount $V=V_{12}-V_{13}$ and the two
electrons are repelled to opposite corners along the diagonals. Although it
is straightforward to diagonalise the Hamiltonian matrix with all 6 base
states (24 including spin), we may obtain the lowest states by perturbation
theory and the system will resonate between the two low-energy base states
with amplitude $\Delta =2t^{2}/V$. This is described by the effective
Hamiltonian (given by degenerate perturbation theory) 
\begin{equation}
H_{{\rm eff}}=\Delta (n_{1}n_{3}-n_{2}n_{4})(R_{\pi /2}-R_{-\pi /2})+J\left[
({\bf s}_{1}\cdot {\bf s}_{3}-1/4)n_{1}n_{3}+({\bf s}_{2}\cdot {\bf s}%
_{4}-1/4)n_{2}n_{4}\right]  \label{two}
\end{equation}
where $R_{\theta }$ is a rotation operator, {\it i.e.} 
\[
R_{\pi /2}|\sigma ,*,\sigma ^{\prime },*\rangle =|*,\sigma ,*,\sigma
^{\prime }\rangle \quad , 
\]
etc. Note that the prefactor ($n_{1}n_{3}-n_{2}n_{4}$) takes the value $\pm
1 $ depending on whether either `sites' 1 and 3, or 2 and 4 are occupied.
This effective Hamiltonian is easily diagonalised. The spectrum consists of
two singlets and two triplets. We see immediately that (\ref{two}) gives
zero when operating on a spin-polarised state and hence the triplets are at
zero energy with the singlets at energies $-J\pm 2\Delta $, in agreement
with what is obtained from the pocket state approximation for this problem
when the exchange of the two electrons is included there \cite{haus94}.
There are thus two contributions to the `binding energy' of the singlet
ground state, a resonance energy ($-2\Delta )$ and a superexchange energy ($%
-J$). It is easy to see that the former will be dominant since it occurs in
second-order and does not involve double-occupation of a `site'. On the
other hand, there are two contributions to $J$, the second-order term $\sim
4t_{13}^{2}/U$ and fourth-order terms $\sim t_{12}^{4}/V^{2}U$.

The cases of four and five electrons on a square reduce directly to an
effective Heisenberg model and the corresponding spectra are shown in
Fig.~12b and 12c. We note that the ground-state of $N=4$ has $S=0$ whereas
for $N=5$ the ground has $S=3/2$, {\it i.e.} not the low-spin ground state.
This is easily understood from the Heisenberg model which, due to the
antiferromagnetic exchange interaction, will tend to align adjacent spins
antiparallel. With 4 electrons this gives $S=0$ but with 5 electrons the
spin on the electron in the centre of the square will tend to align
antiparallel with those in the four corners, giving $S=3/2$. (Note that the
exchange interaction between electrons on neighbouring corners will oppose
this tendency to parallel alignment but, since the distance is a factor $%
\sqrt{2}$ greater than to the centre, this `frustration' is insufficient to
enforce low-spin.)

This difference in the nature of the ground states for 4 and 5 electrons in
a square dot has important consequences for single-electron transport. This
is shown in Fig. 14 where we consider a situation where the ground state
energy of the dot with 5 electrons is exactly degenerate with that of 4
electrons in their ground state and one electron at the Fermi energy in a
lead. (We assume that the source-drain voltage is vanishingly small.)
Strictly speaking, these states cannot be connected by a tunnelling
interaction at $T=0$ since this is forbidden by conservation of spin, {\it %
i.e.} to be allowed the total spin $S$ on the dot may only differ by $\pm 1/2
$ when the electron number fluctuates by one. This phenomena is known as 
{\em spin blockade} \cite{spinblock} since, if we imagine starting in the
Coulomb blockade regime with 4 electrons on the dot in their ground-state
with energy $E_{4}<E_{5}-E_{F}$ (where $E_{5}$ is the ground-state energy
with 5 electrons), and raise the energy of $E_{4}$ (via the gate voltage),
then the current through the dot will be spin-blocked even when $%
E_{4}+E_{F}=E_{5}$. By further raising the energy of the electrons on the
dot relative to the Fermi energy, the single-electron current will
eventually be `switched on' when the $S=1/2$ states with 5 electrons become
degenerate with $E_{4}+E_{F}$. A finite conductance may also be recovered
from the spin-blockade regime by either raising the temperature, when
excited states with $S=1/2$ again become involved, or by increasing the
source-drain voltage, which has a similar effect. These effects are generic
in the sense that they will apply to a variety of quantum-dot geometries
(either contrived or accidental). In this way the `recovery' of `missing'
conductance peaks when heating the sample up to 500 mK, as observed in \cite
{temprec}, can be interpreted by such a spin blockade phenomenon \cite
{annals}.

Although the above equivalence of a few-electron quantum-dot system to
Hubbard/charge-spin models is adequate for most situations, there is one
further class of problem for which it breaks down. These are problems with
very high symmetry in which processes involving the simultaneous movement of
more than two electrons can be important or even dominant in comparison to
processes where just two neighbouring electrons are exchanged. A simple
example would be an (almost) circular dot that contains a very small number (%
$N=2\ldots 6$) of electrons and is sufficiently large that the mean
separation between electrons exceeds the Bohr radius. In the classical
Wigner limit there are low-energy states in which all $N$ electrons arrange
themselves around the periphery of the dot and rotate whilst preserving
their relative separations. This indicates that the corresponding elementary
permutational process, a cyclic exchange of all electron places, is
connected with a larger probability amplitude than the exchange of two
nearest neighbours. This affects considerably the spin multiplets at low
energies. For example, an odd electron number favours a high-spin
ground-state while $S=0$ for $N$ even. Another example where `higher order'
permutations processes could be important is inside the bulk of an extended
Wigner crystal (or, a large dot containing many electrons at low-density)
where the electrons arrange themselves in the form of an (almost) hexagonal
lattice. There, $2\pi /3$ rotations of any of the electron triangles
composing the lattice could well be of similar importance to the pair
exchange of adjacent electrons.

Can the Hubbard/charge-spin models be modified to accommodate these
processes? The answer is yes and in fact similar processes can also be
important for lattice systems, though this is not well-kown. An example of
the latter is a contribution to the superexchange between two holes in the
copper-oxide planes of the high temperature superconductors in which the
holes never occupy the same sites simultaneously.\cite{eskesjeff} For the
quantum dot, we may illustrate these processes by considering a triangular
quantum dot. By introducing intermediate lattice points, we can rotate all 3
electron through $2\pi /3$ in a series of 6 steps as shown in Fig. 15. The
intermediate steps are higher in energy than the initial and final states
(the electrostatic Coulomb energy is higher) and may be eliminated by
perturbation theory. The result is an effective interaction of the form $%
-K(R_{2\pi /3}+R_{-2\pi /3})$, where $K$ is a constant energy and $R_{2\pi
/3}$ ($R_{-2\pi /3}$) is an operator which performs a cyclic (anticyclic)
rotation of all three electrons. We note that these rotation operators may
be written as spin-operators using the identities, 
\[
R_{2\pi /3}\equiv P_{12}P_{23}{\rm ~and~}R_{-2\pi /3}\equiv P_{23}P_{12} 
\]
where $P_{ij}\equiv 2{\bf s}_{1}\cdot {\bf s}_{2}+1/2$ is the Dirac exchange
operator. Hence, expressing the exchange term in Eq. (\ref{tJVmodel}) in
terms of these permutation operators and dropping an overall constant
energy, the effective spin-Hamiltonian for 3 electrons in an equilateral
triangular potential well becomes, 
\[
H_{{\rm eff}}=\frac{J}{2}\left[ (P_{12}-1)+(P_{23}-1)+(P_{31}-1)\right]
-K(P_{12}P_{23}+P_{23}P_{12})\quad. 
\]
We see immediately from this form that the spin polarised state has
eigenenergy $-2K$. It is straightforward to complete the diagonalisation of $%
H_{{\rm eff}}$ which results in a further pair of degenerate doublets at
energy $K-3J/2$. We refer to the $K$-term as a `ring' term and we see that
it opposes the exchange ($J$) terms, with the former favouring a high--spin
ground state and the latter a low--spin ground state. The crossover occurs
at $J=2K$, which is not quite reached for the equilateral triangle, as
discussed above.

This will not always be so for other geometries as follows immediately from
the fact that we can change the shape of the dot in such a way that the ring
processes are increased in amplitude relative to the exchange processes, as
shown, for example, in Fig. 16b. An `extreme' example of this is the
circular dot (Fig. 16c) which, as discussed earlier, has a spin polarised
ground-state ($S=3/2$). The transition from high-spin to low-spin ground
states in rings with increasing impurity barrier has been discussed recently
in the context of persistent currents.\cite{trieste}

In summary, we have shown in this section how the low-density few-electron
problem in a quantum dot may be mapped onto a Hubbard model and that this
may be further reduced to a charge-spin model. In one dimension we always
get an effective Heisenberg model with antiferromagnetic exchange and a
low-spin ground state. In higher dimensions there are other possibilities
which depend upon dot geometry and the number of electrons. This can have a
dramatic effect on the nature of the energy spectrum, both for the ordering
of the states and the energy scale of the separations between levels. For
geometries in which there is only one electrostatic configuration in the
classical limit, the degeneracy due to the permutation of electrons on
different `lattice sites' is partly lifted by quantum mechanical exchange
described by an antiferromagnetic Heisenberg model. In this cases there is
just one energy scale for the splitting between multiplets, the exchange
energy $J$. Note, however that in more than one dimension the ground state
spin can be larger than minimum, an example being 5 electrons in a square.
In other situations, where there is more than one classical electrostatic
configuration of lowest energy, the energy separations of multiplets may be
quite different. For example, for 3 electrons (1 hole) in a square there
will be energy splittings on both the scale of $t$ and of $J$, with $t\gg J$
and, furthermore, the ground-state will be high spin. The so-called `ring'
processes can also be important. These effectively represent the
simultaneous cyclic permutation of more than two electrons and have a
characteristic energy scale $K$. For the example of 3 electrons in a
triangle, $K$ approaches the rotational constant $h^{2}/6mL^{2}$ of a
perfect circular dot of circumference $L$, as the triangle is distorted into
a circle, when there will be energy spilittings on the scale of both $K$ and 
$J$, with a crossover from low-spin to high-spin ground state. More
complicated situations can be envisaged as the number of electrons is
increased. Thus, for example, 7 electrons in a square dot will have a
classical ground-state configuration with four electrons in the corners of
the dot and the remaining three forming a triangle in the interior of the
dot. Important quantum mechanical processes will include pair exchange and
`ring' processes in which the three interior electrons will perform `rigid'
rotations of $\pi /6$. Future work could investigate these processes in more
detail and estimate their relative magnitudes using quasi-classical methods,
in addition to investigating other geometries which show promise for device
applications.

\section{Future prospects}

The science of quantum dots and artificial atoms is still very much in its
infancy and we can expect many new developments in the forthcoming years as
technological advances permit reliable and reproducible fabrication of
device structures. From the present perspective, a number of areas with both
fundamental and applications potential seem worthy of further investigation.

Single dots which are not too small offer the possibility of investigating
in a controlled way both the Wigner `lattice' and the transition from the
Wigner regime to the independent electron (quasiparticle) regime as the
electron number and mean electron density is increased. As disussed in the
previous section, the fundamental difference between the Wigner regime in a
`few-electron' quantum dot and the infinte Wigner lattice is the major
effect of the boundaries that can increase the ground state spin to values
greater than $S=0$ or $S=1/2$, and provides a separation of the energy
scales related to `vibrational' and spin degrees of freedom, respectively.
In principle, this opens two possible scenarios towards the Wigner
transition. Firstly, we can increase the system size whilst maintaining a
sufficiently low but constant density to bring the excited vibrational
states closer to the ground manifold. They eventually will mix with the spin
states which changes the low energy physics qualitatively. On the other
hand, we can imagine to increasing the confinement with a constant number of
electrons. This will suppress the (effective) spin dependence of the low
energy states and leave the quantum dot with classical electrons, in analogy
to the transition from the Heitler-London to the molecular orbital picture
in molecules. This crossover is also related to the metal-insulator (Mott)
transition which can take place in strongly correlated lattice systems.

Another area worthy of further investigation is the magnetic field
dependence in the Wigner regime. Large magnetic fields modify the low energy
eigenstates considerably \cite{fock28}, as it is observed also
experimentally in quantum dots \cite{mceuen91,ashoori93a,weis94,zhitenenv}.
However, as a pure quantum effect even a non-vanishing vector potential
(which does not require a magnetic field at the location of the electrons at
all) can have very interesting consequences. One famous example is that of
finite currents circulating around a small ring due to the Aharonov--Bohm
flux $\phi $ enclosed by the ring (which is an equilibrium effect, in
contrast to superconducting currents~!). It is known that electron--electron
interactions are important, otherwise the experimentally observed magnitudes
of the currents cannot be understood. Theoretically, this problem can be
approached by considering a one-dimensional quantum dot, such as that
described in Section~\ref{theory}, deforming it to form a ring. The electron
molecule can then rotate (almost) freely, maintaining the distances between
the electrons. The ground state energy $E_{0}$ and, most importantly, the
ground state spin $S$, can then be determined in a similar way to that
described in Section~\ref{theory} \cite{trieste}. However, they depend now
on $\phi $. This determines the persistent current that is proportional to
the derivative $\partial E_{0}(\phi )/\partial \phi $ (at zero temperature).
Since even impurities along the ring do not cause level repulsion between
energies belonging to different spin values $S$ and $S^{\prime }$, the
persistent current can exhibit discontinuous jumps near values for the flux
where level crossings occur. In disordered rings this would not occur for
spinless electrons. The magnitude of the persistent current indeed increases
with increasing interaction strength. The main shortcoming of this model of
a very thin, one-channel ring is that it cannot explain values of the
current observed experimentally, which correspond to almost freely
circulating electrons. \cite{pcexp}. However, it has became clear that
maximum persistent currents in disordered rings require the investiation of
wider, many channel rings. For any description within the picture of Wigner
localized electrons this means that we need to increase the electron number
in the ring considerably. This is only feasible using an approach that is
highly selective in the Hilbert space describing the low-energy physics. The
spin-charge representation described in Section~\ref{theory} should provide
a means of achieving this.

There are many new effects to be investigated in multiple quantum dots and
quantum dots arrays, {\it i.e.} artificial molecules and solids. Unlike
their `real' counterparts, they have the prospect of being controlled in a
well defined way as technology advances. Thus, for example, we have the
prospect of studying the classic Mott transition \cite{mott} for an array of
one-electron artificial atoms as the ratio of the tunnelling matrix element
between dots to the Coulomb charging energy is varied. Another way of
inducing a metal-insulator transition is to change the number of electrons
on each artificial atom by simply varying the chemical potential through
gate voltages. The insulator to metal transition occurs when the Fermi
energy exceeds the minimum tunnelling barrier between dots and a dot array
becomes an `antidot' array.\cite{antidot} The `insulating' regime of such an
array of one-electron artificial atoms would be interesting in its own right
since, as with real atoms, we might expect an antiferromagnetic
superexchange interaction between neighbouring dots. With more electrons/dot
we could `engineer' artificial atoms with high-spin ground states (by choice
of dot geometry and electron number). These would give rise to magnetic
correlations between dots and could, in principle introduce magnetic low
energy modes and, eventually, magnetically ordered phases at sufficiently
low temperatures. More generally, in the long-term we can invisage
`designer' molecules and solids with desired properties (energy structure,
band structure, conductivity, susceptibility, anisotropy etc.)!

Finally, we may speculate on the feasibility of practical applications of
quantum dots, single-electron transistors etc. The main obstacle to their
widespread use for electronic applications is the need for low-temperatures.
The temperature scale is ultimately set by the size of the quantum dot and,
in principle, there is no fundamental reason why devices should not be made
sufficiently small to operate at room temperature or higher. Indeed, there
has been a great deal of progress over the past five years during which we
have seen the temperature of operation of single-electron transistors
increase from 4K to 77K, with evidence for oscillations in conductance due
to single-electron tunnelling being discernible even at room temperature.%
\cite{roomtemp} Similar evidence for room-temperature, single-electron
memory devices has also been reported recently.\cite{memory} There is little
doubt that this progess will continue with the relentless drive towards
smaller structures as nanoelectronics approaches the molecular level. With
the need for ever increasing levels of integration and low-power consumption
there is also little doubt as to the potential markets for single
electronics, though the formidable technological problems associated with
reproducibility and fabrication on the nanoscale should not be
underestimated. Other promising areas for quantum dot arrays and quantum
wires are optical applications, particularly low-threshold lasers. Here
again the key issues at present are technological, with the need to
reproduce devices with a small variance in size. Despite these difficulties,
recent progress is encouraging and prospects are good for the use of quantum
dots and artificial atoms to investigate fundamental physics and the
development of new device concepts into the twenty-first century.\medskip

{\bf Acknowledgements\medskip }

We acknowledge valuable discussions with Colin Lambert, Hermann Grabert,
Sarben Sarkar and colleagues in our EU-sponsored HCM network on the quantum
dynamics of phase coherent structures (HCM No. CHRX-CT93-0136). JHJ thanks
the organizers of the 20th International School on Theoretical Physics for
their invitation to participate and kind hospitality in Ustron. WH also
wishes to thank the University of Minnesota for warm hospitality.

\newpage {\bf Figure Captions}

\medskip

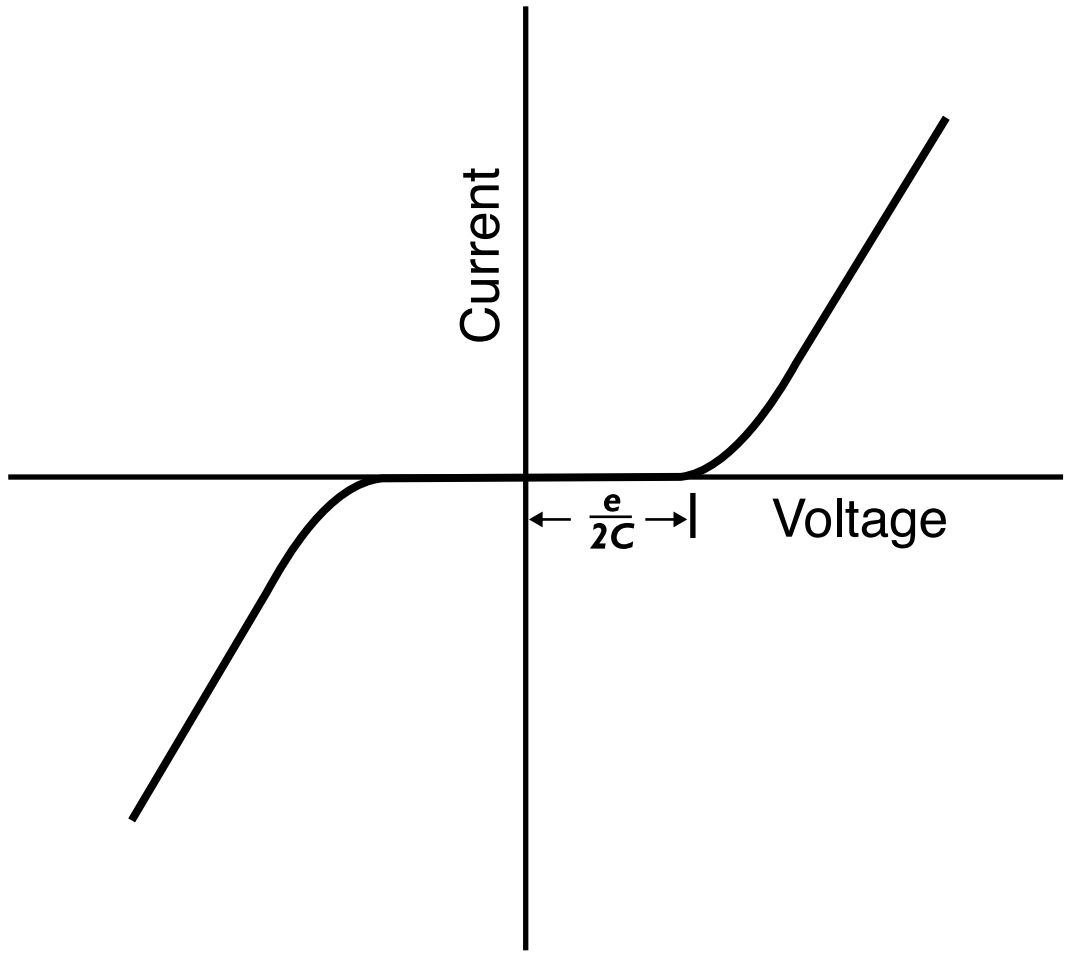
\noindent FIG. 1. Current {\it vs } voltage for a small metal particle,
showing the `Coulomb gap'.

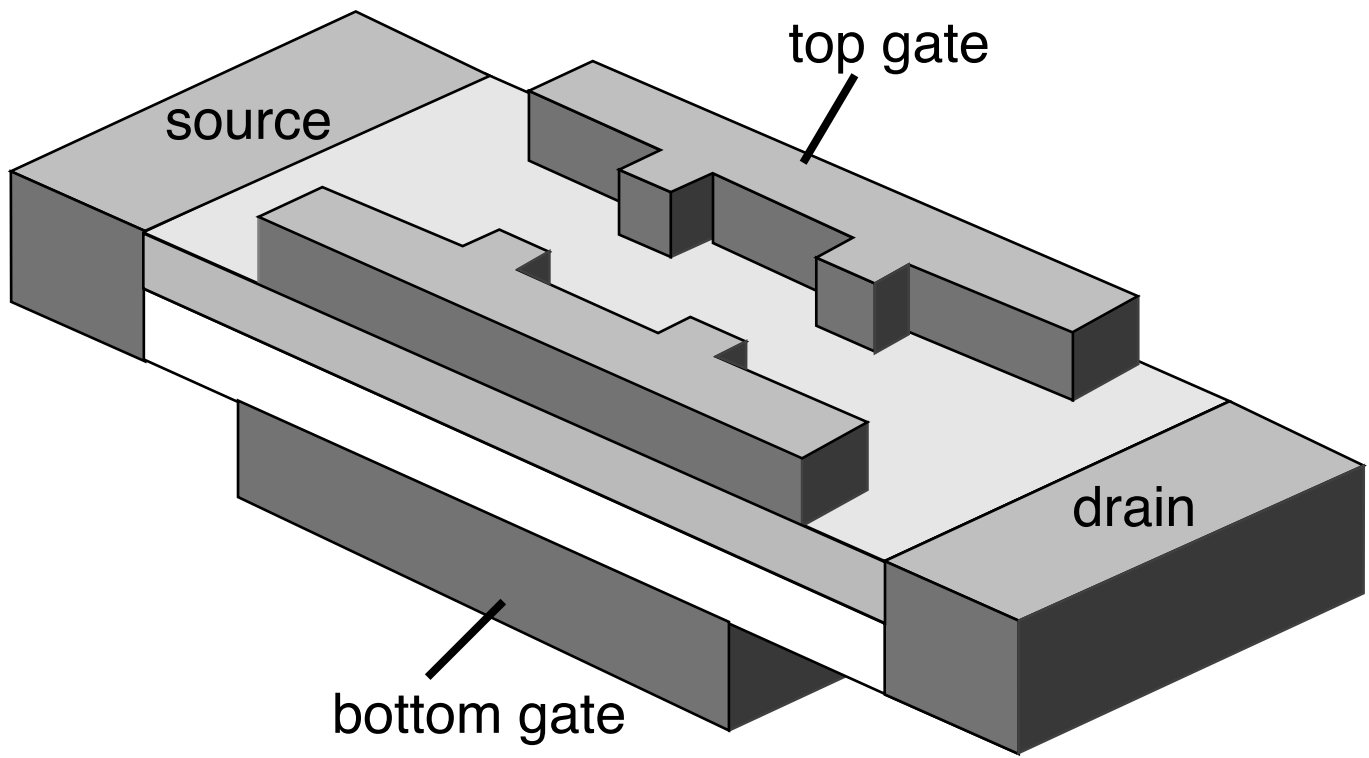
\noindent FIG. 2. Schematic diagram of a single-electron transistor.

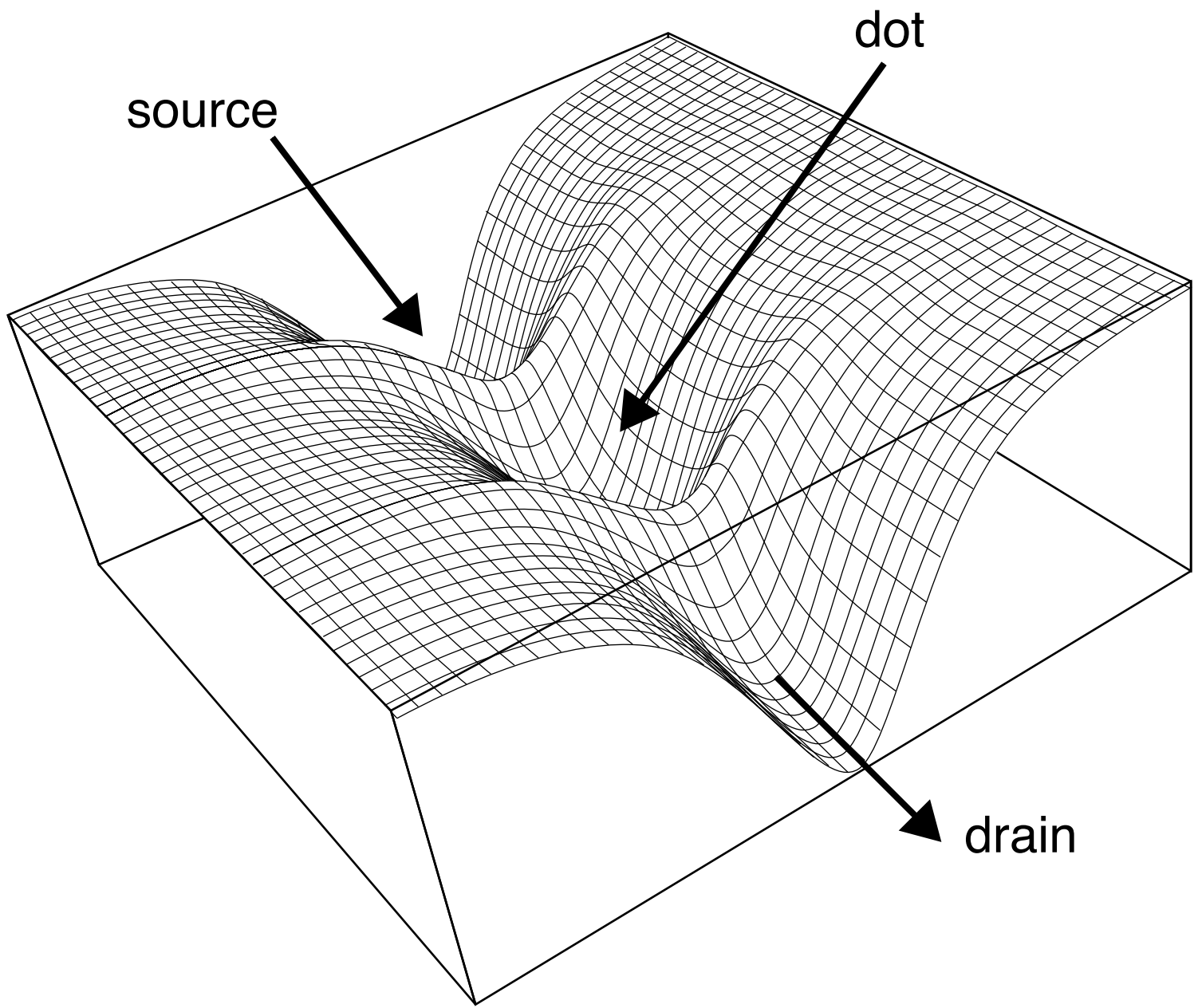
\noindent FIG. 3. Electron potential-energy landscape in the plane of the
two-dimensional electron sheet of the single-electron transistor of Fig. 2.

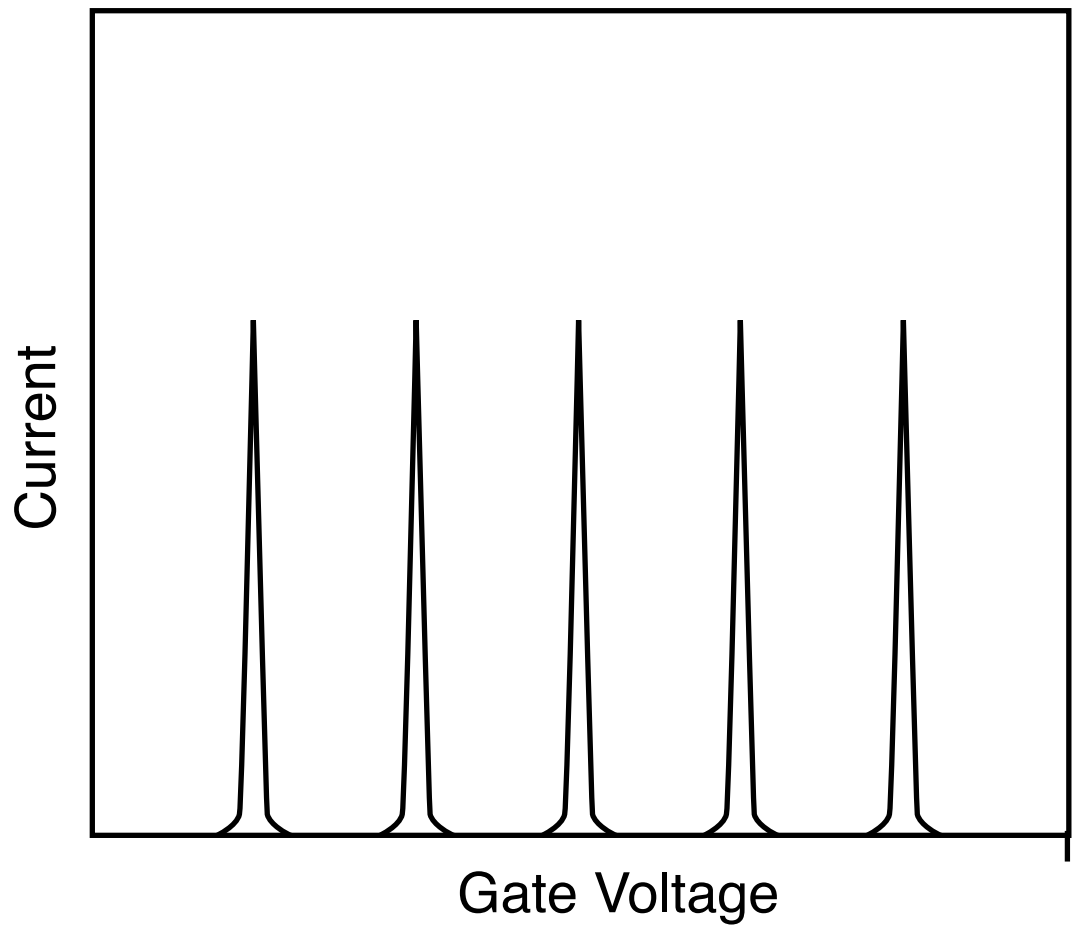
\noindent FIG. 4. Current {\it vs} gate voltage for a single-electron
transistor with small souce-drain bias, showing single-electron Coulomb
oscillations.

\noindent FIG. 5. Potential landscape for the single-electron transistor
showing (a) empty quantum well with no current at $T=0$ and (b) quantum well
occupied with finite current at $T=0$.

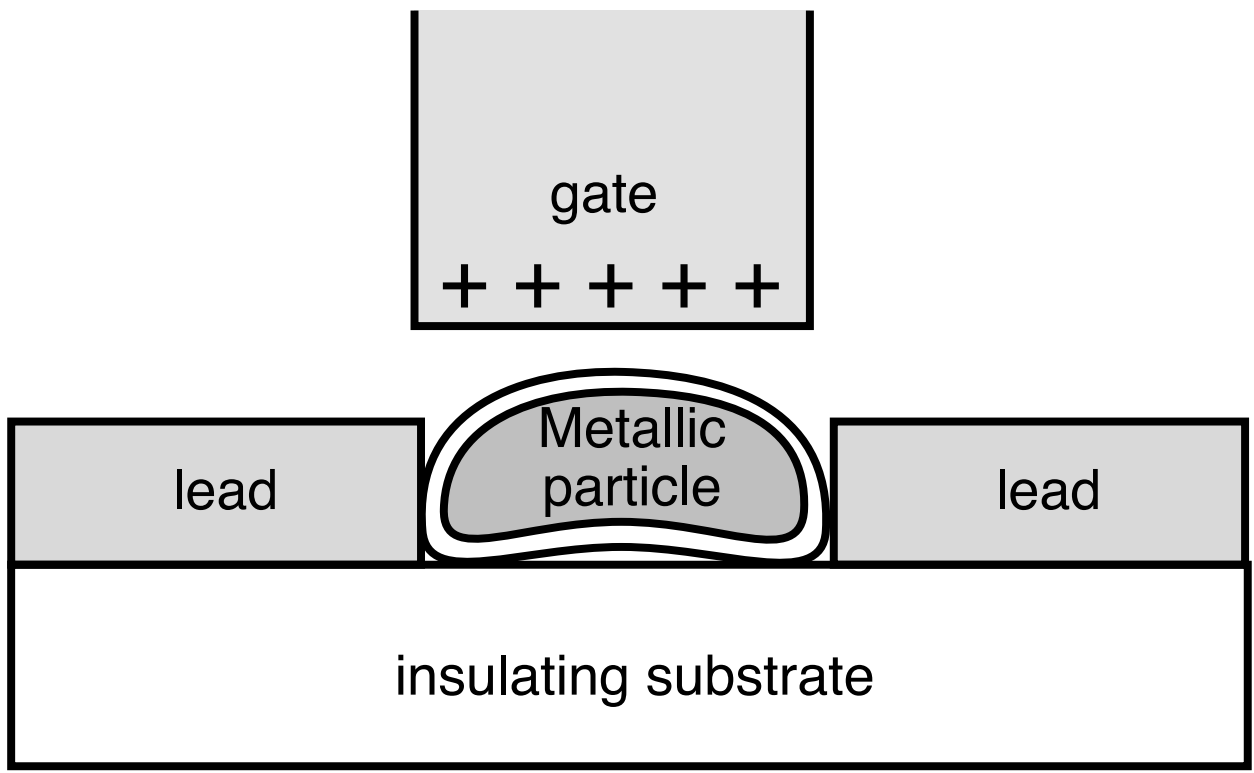
\noindent FIG. 6. Schematic diagram of small metal particle on an insulating
substrate with source, drain and gate leads.

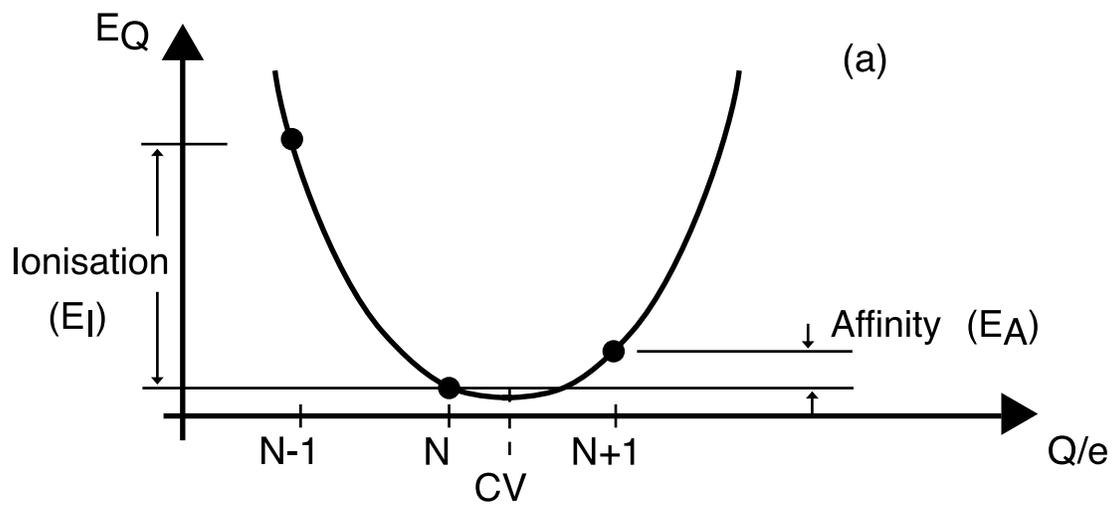
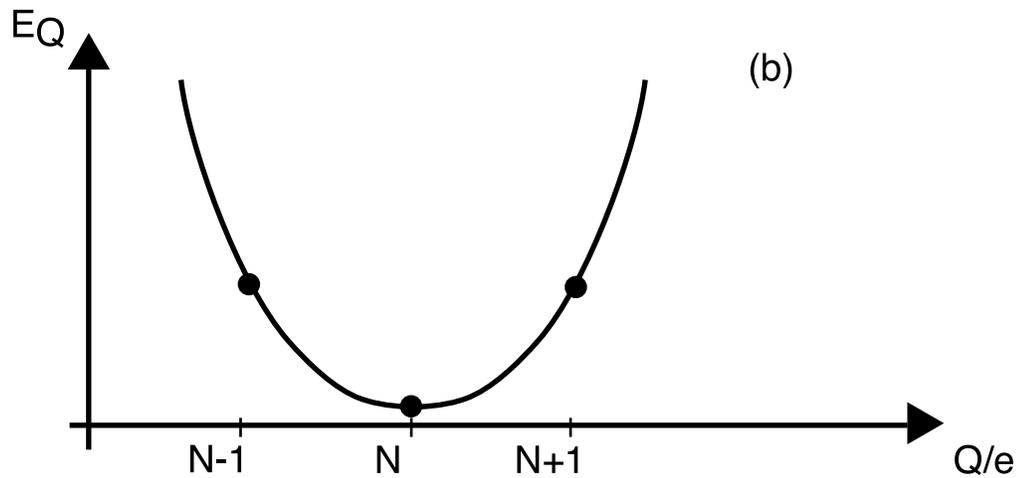
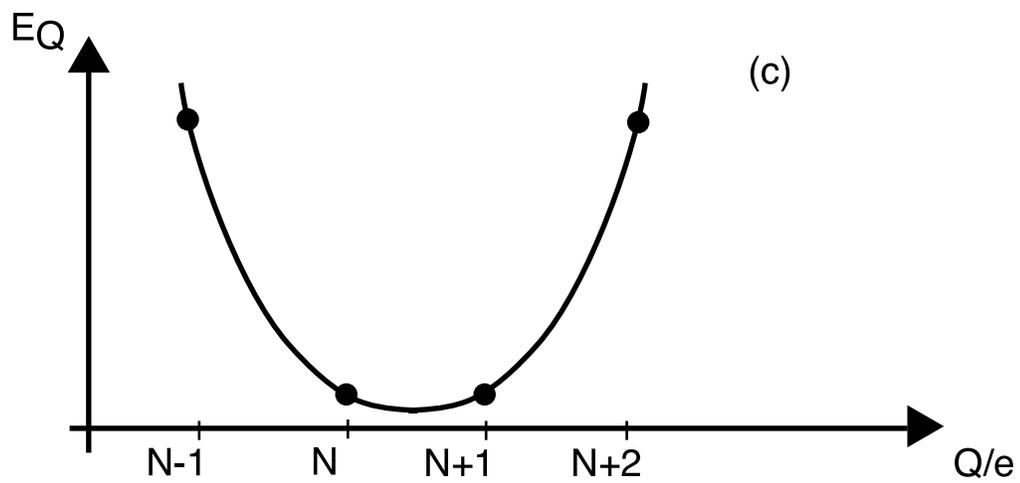
\noindent FIG. 7. Energy parabolas for the semi-classical charging model.
The bullets indicate the possible values for the metallic island. (a)
arbitrary gate voltage showing quasi-atomic ionisation energy ($E_{I}$) and
electron affinity ($E_{A}$). (b) situation mid-way between two successive
conductance peaks (c) situation right at a conductance peak.

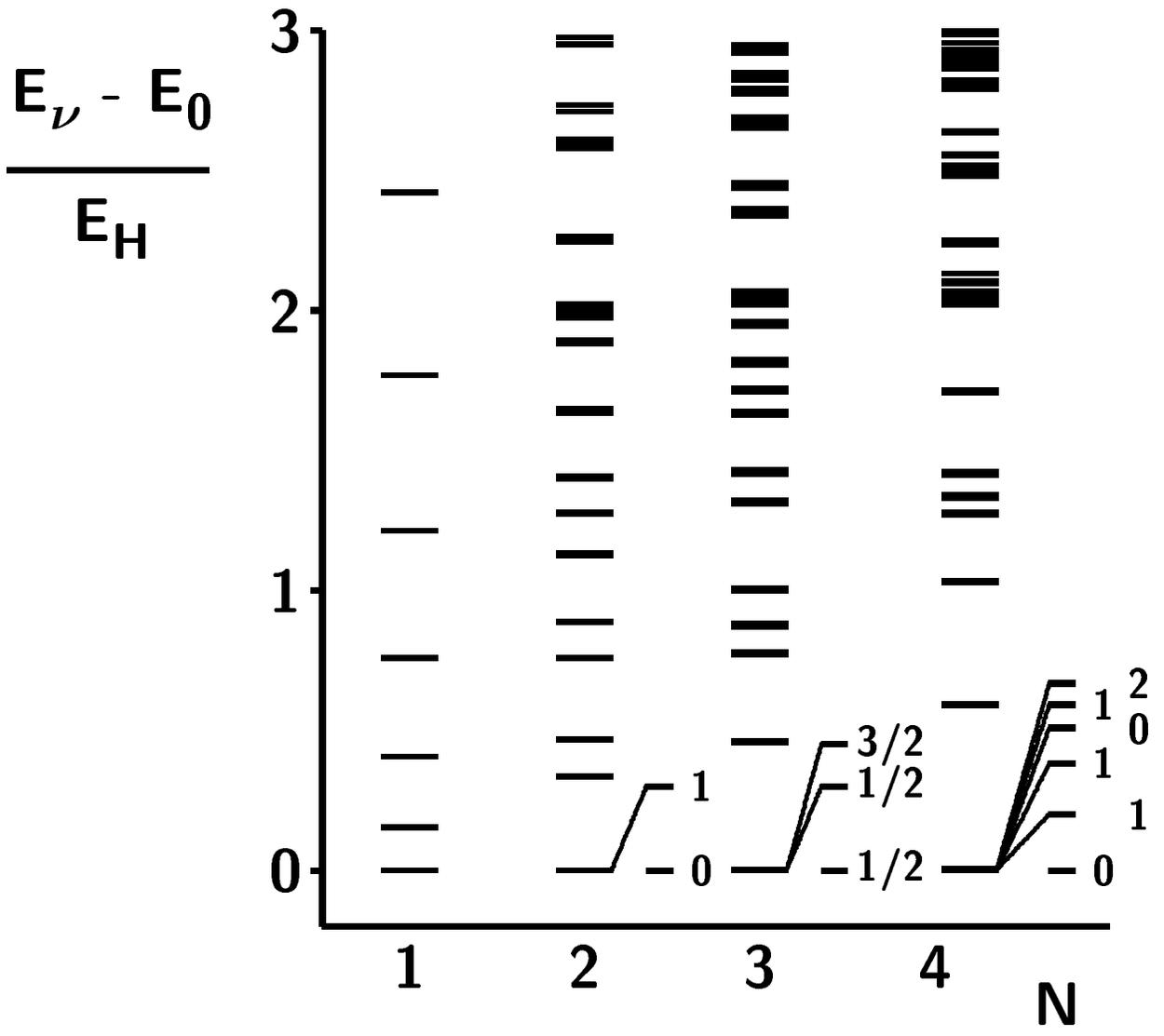
\noindent FIG. 8. Typical spectra of square well model in 1D for $N=1,\ldots
,4$ and $L=9.45a_{{\rm B}}$. For $N\ge 2$ the low-lying eigenvalues form
groups of (fine structure) multiplets, the total number of states per
multiplet being equal to the dimensionality of the spin Hilbert space, $2^{N}
$. For clarity the lowest multiplets are magnified indicating the total spin
of each level. The ground state energies are subtracted.

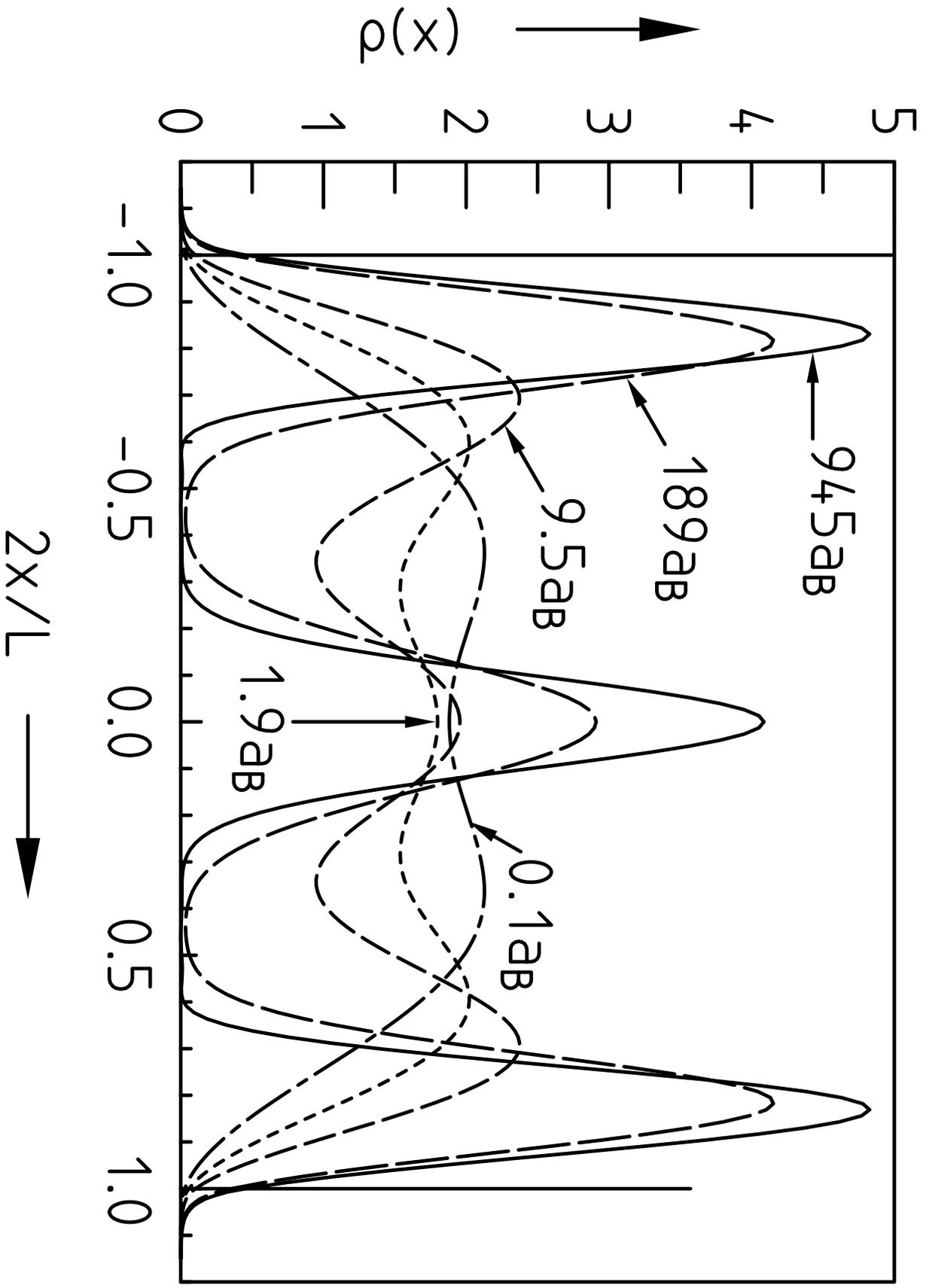
\noindent FIG. 9. Charge density $\varrho (x)$ for three electrons and
various $L$, with normalization such that $\int \varrho (x)dx=3$. When $%
L\gtrapprox a_{{\rm B}}$, three peaks begin to emerge and become well
separated for $L\gtrapprox 100a_{{\rm B}}$.

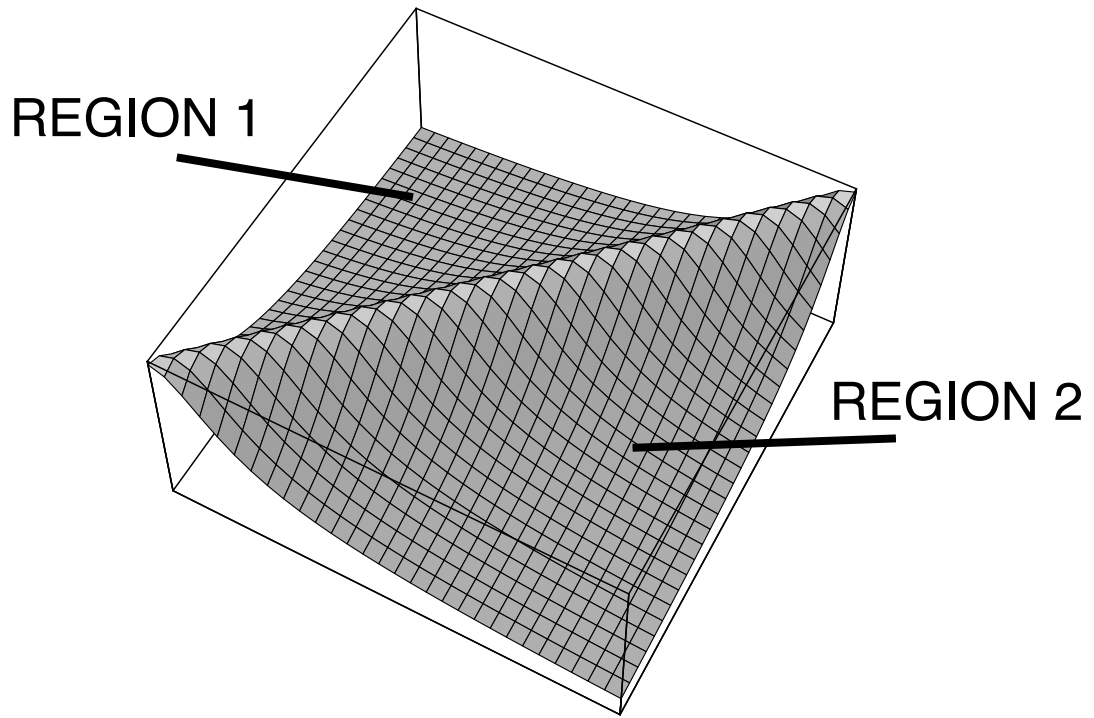
\noindent FIG. 10. Potential landscape for two electrons in a 1D well,
equivalent to the potential seen by a single (fictitious) equivalent
particle in 2D.

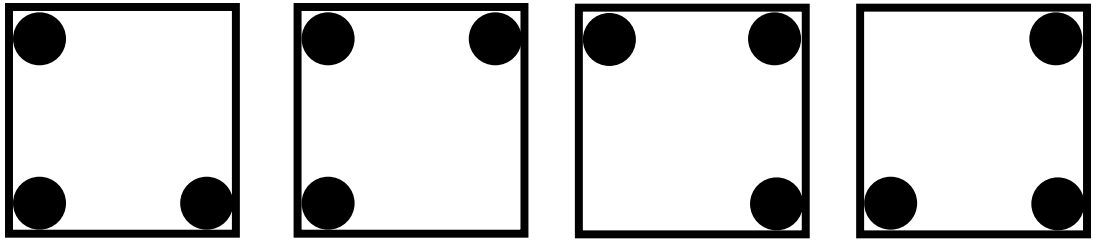
\noindent FIG. 11. Classical configurations with minimum electrostatic
energy for three electrons in a square dot. Quantum mechanically the system
may be described approximately by a tight-binding model ($t$-model) in which
the `hole' hops between adjacent corners.

\noindent FIG. 12. Low-energy states of of interacting electrons in a square
quantum dot. (a) 3 electrons given by solutions of the $t$-model, showing a
spin-polarised ground state. (b) 4 electrons with singlet ground state from
solution of the antiferromagnetic Heisenberg model. (c) 5 electrons with $%
S=3/2$ ground state, again from the antiferromagnetic Heisenberg model.

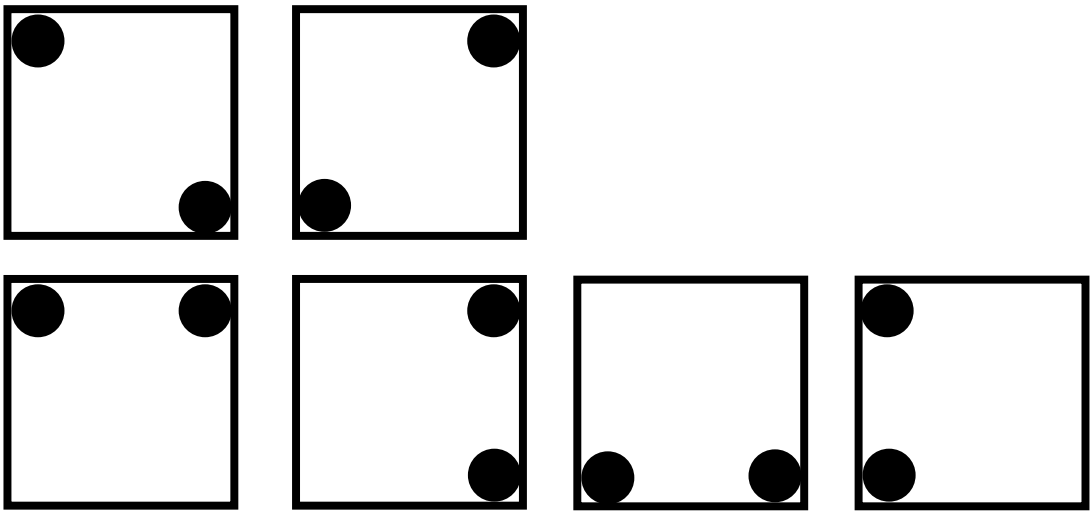
\noindent FIG. 13. Low-energy classical configurations for two electrons in
a square quantum dot which are described quantum mechanically by the $tJV$%
-model. The last four (higher energy) configurations may be eliminated by
degenerate perturbation theory, resulting in a spin model which resonates
between the two lowest-energy configurations.

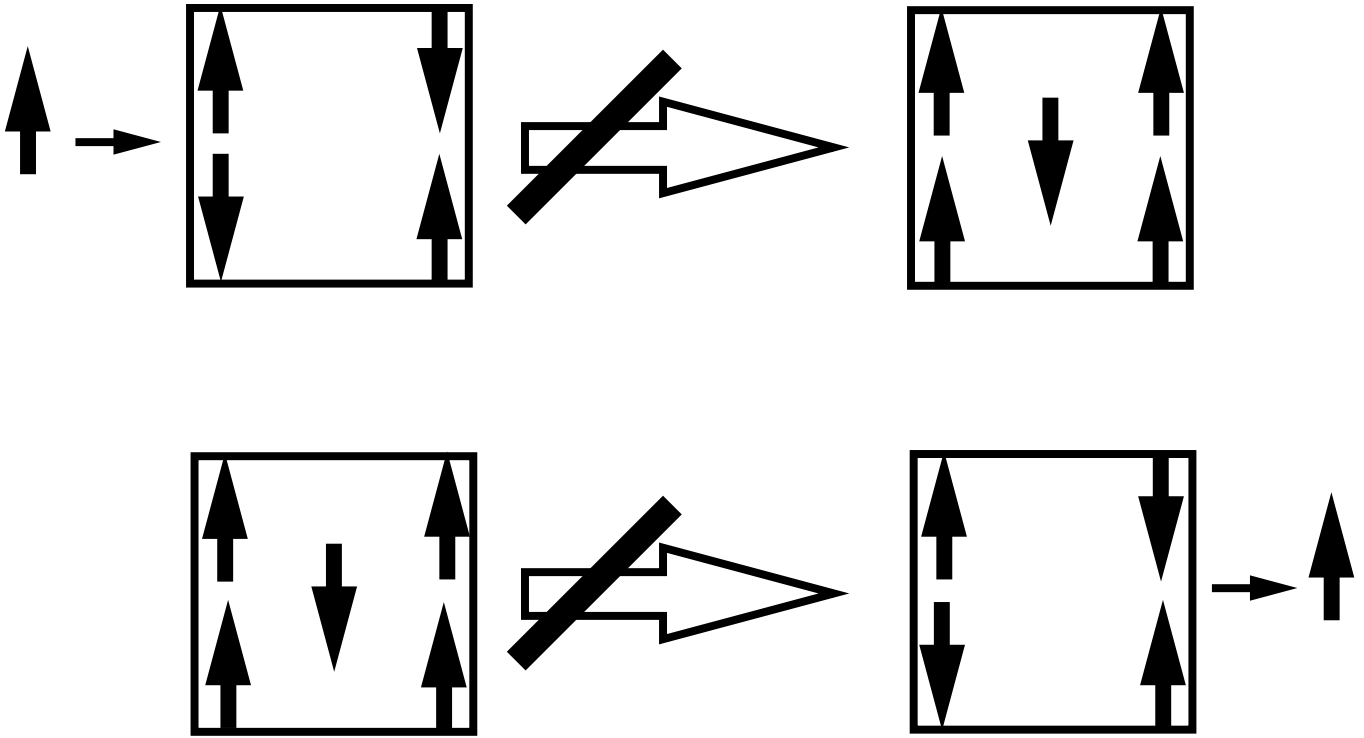
\noindent FIG. 14. Electron configurations showing spin-blockade. Although
the ground-state with four-electrons in a square dot plus an electron at the
Fermi energy is degenerate with the ground-state with five electrons on the
dot (so energy conservation would allow the current to flow), the transition
is forbidden since the electron entering is not able to change the
ground-state spin on the dot from $S=0$ ($N=4$) to $S=3/2$ ($N=5$). The same
is true for the transition with one electron leaving the dot with five
electrons.

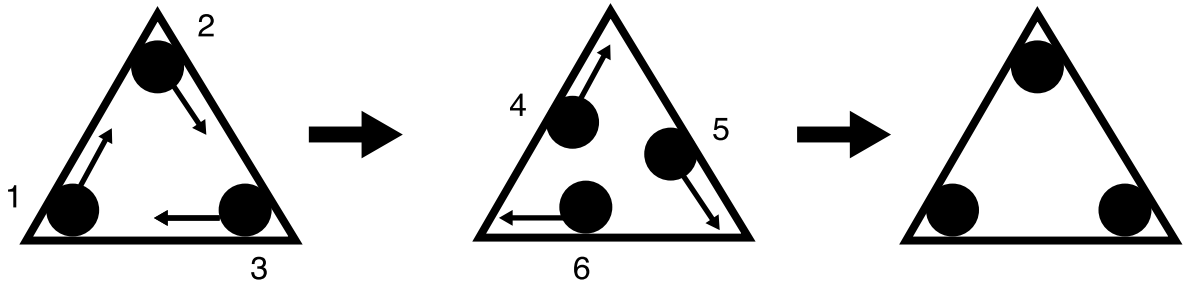
\noindent FIG. 15. A six-step ring process for a triangular quantum dot, in
which all three electrons rotate cyclically through $2\pi /3$.

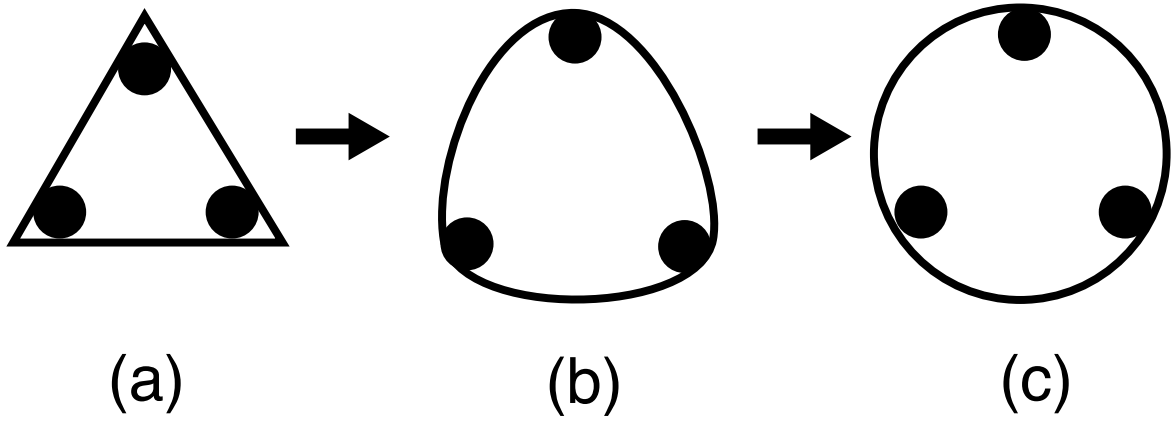
\noindent FIG. 16. Gradual distortion of a triangular quantum dot into a
circular dot during which the ground state switches from $S=1/2$ (a) to $%
S=3/2$ (c).

\newpage\vspace*{\fill}\hspace*{-2cm}\includegraphics{f1}

FIGURE 1 \hspace{1cm} JEFFERSON/H\"AUSLER

\newpage\vspace*{\fill}\hspace*{-2cm}\includegraphics{f2}

FIGURE 2 \hspace{1cm} JEFFERSON/H\"AUSLER

\newpage\vspace*{\fill}\hspace*{-2cm}\includegraphics{f3}

FIGURE 3 \hspace{1cm} JEFFERSON/H\"AUSLER

\newpage\vspace*{\fill}\hspace*{-2cm}\includegraphics{f4}

FIGURE 4 \hspace{1cm} JEFFERSON/H\"AUSLER

\newpage\vspace*{\fill}\hspace*{-2cm}\includegraphics{f5a}

FIGURE 5a \hspace{1cm} JEFFERSON/H\"AUSLER

\newpage\vspace*{\fill}\hspace*{-2cm}\includegraphics{f5b}

FIGURE 5b \hspace{1cm} JEFFERSON/H\"AUSLER

\newpage\vspace*{\fill}\hspace*{-2cm}\includegraphics{f6}

FIGURE 6 \hspace{1cm} JEFFERSON/H\"AUSLER

\newpage\vspace*{\fill}\hspace*{-2cm}\includegraphics{f7}

FIGURE 7 \hspace{1cm} JEFFERSON/H\"AUSLER

\newpage\vspace*{\fill}\hspace*{-2cm}\includegraphics{f8}

FIGURE 8 \hspace{1cm} JEFFERSON/H\"AUSLER

\newpage\vspace*{\fill}\hspace*{-2cm}\includegraphics{f9}

FIGURE 9 \hspace{1cm} JEFFERSON/H\"AUSLER

\newpage\vspace*{\fill}\hspace*{-2cm}\includegraphics{f10}

FIGURE 10 \hspace{1cm} JEFFERSON/H\"AUSLER

\newpage\vspace*{\fill}\hspace*{-2cm}\includegraphics{f11}

FIGURE 11 \hspace{1cm} JEFFERSON/H\"AUSLER

\newpage\vspace*{\fill}\hspace*{-2cm}\includegraphics{f12}

FIGURE 12 \hspace{1cm} JEFFERSON/H\"AUSLER

\newpage\vspace*{\fill}\hspace*{-2cm}\includegraphics{f13}

FIGURE 13 \hspace{1cm} JEFFERSON/H\"AUSLER

\newpage\vspace*{\fill}\hspace*{-2cm}\includegraphics{f14}

FIGURE 14 \hspace{1cm} JEFFERSON/H\"AUSLER

\newpage\vspace*{\fill}\hspace*{-2cm}\includegraphics{f15}

FIGURE 15 \hspace{1cm} JEFFERSON/H\"AUSLER

\newpage\vspace*{\fill}\hspace*{-2cm}\includegraphics{f16}

FIGURE 16 \hspace{1cm} JEFFERSON/H\"AUSLER

\end{document}